\documentclass[a4paper,11pt]{article}
                                                                                                                                                                                                                                                                                                                                                                                                                                                                                                                                                           
\usepackage{amssymb,amsthm,amscd, amsbsy, array,amsmath,mathdots}
\usepackage{cancel}
\usepackage[francais]{babel}
\usepackage{fontenc}
\usepackage{amsfonts}
\usepackage[mathscr]{eucal}
\usepackage{mathrsfs}
\usepackage{latexsym}
\usepackage{graphicx}
\usepackage[a4paper,textwidth=17cm,textheight=21cm]{geometry}

\newcommand{\be}{\begin{equation}}
\newcommand{\ee}{\end{equation}}
\newcommand{\bea}{\begin{eqnarray}}
\newcommand{\eea}{\end{eqnarray}}

\def\1{{\mbox{\boldmath $1$}}} %
\newcommand{\C}{\mathbb{C}}

\newtheorem{proposition}{Proposition}

\newtheorem{lemme}{Lemma}
\newtheorem{theorem}{Theorem}

\begin{document}
\thispagestyle{empty}
\begin{center}
	\begin{huge}
		Noncommutative ricci curvature and dirac operator on $B_q[SU_2]$ at the fourth root of unity
	\end{huge}
\end{center}

\begin{center}
	\begin{large}
		ARM Boris Nima \\
	\end{large}
	{\footnotesize  \textit{arm.boris@gmail.com}}
\end{center}

\vspace{5cm}
\begin{center}
	\textbf{Abstract} \\
We calculate the torsion free spin connection  on the quantum group $B_q[SU_2]$\\
 at the fourth root of unity. From this we deduce the covariant derivative\\
  and the Riemann curvature. Next we compute the Dirac 
  operator \\
   of this quantum 
   group and we give numerical \\
    approximations of its eigenvalues.
\end{center}

\newpage

\setlength{\textheight}{21cm}
\addtolength{\voffset}{0.2cm}

\section{introduction}
The quantum group $B_q[SU_2]$  and the 4-dimensional differential calculus with left basis 1-form are defined in \cite{fuzzy}. The goal of this article is to give the expression of the Dirac operator and of its eigenvalues. 
\paragraph{}
First in the preliminaries, we give the description of the quantum group $B_q[SU_2]$ and the relation between its elements. We work here at the fourth root of unity which means that every fourth power of the elements of $B_q[SU_2]$ are 0 or 1.
\paragraph{}
Next we describe the right multiplication and the right coaction. 
Also, we define the Grassman variables (or 1-forms) and relations between its wedge products. Furthermore, we describe the relation between function and the Grassman variables. 
\paragraph{}
As a consequance of the definition of the Maurer-Cartan form of this quantum group, we deduce the Maurer-Cartan equation. In addition we find the commutation relation  between Grassman variables and each product of elements of $B_q[SU_2]$. By the way we calculate the derivative of each product of functions. In any case, we find the projection on the Grassman variables of any function. Then we can compute the left and right coactions.
\paragraph{}
In fact there a Killing on this quantum which give the torsion and cotorsion equations. In other words, with the expression of the left and right coactions, we can deduce the spin connection in solving those equations. However, again with the expression of the left and right coaction, we can deduce the covariant derivative. Besides, with the expression of the covariant derivative, we find the Riemann curvature. 
\paragraph{}
Moreover, with the expression of the spin connection and the antipode of this quantum group, we calculate the Dirac operator which is not similar to the one find by \cite{riemanngeometry}. With the expression of each partial derivative in a well chosen basis from the expression of the right translation operator, we were able to calculate numerically the eigenvalues of the Dirac operator found previously.
\paragraph{}
This work is a copy of \cite{CQSL2} which is done for $\C_q[SL_2]$ instead of $B_q[SU_2]$.

\newpage
\section{Preliminaries}
\paragraph{}
Here we fix notations in the conventions that we will use and do some preliminary computations, in Section 2.2. We let $q^2 \neq 1$. The quantum group $\mathcal{A}=B_q[Su_2]$ has a matrix of generators $u=
\left(\begin{array}{cc}
\alpha & \beta\\
\beta^* &\delta \\
   \end{array}
\right)$
(see \cite{fuzzy}) with relations 
\begin{equation}
\beta \alpha =q^2 \alpha \beta, \hspace{3mm} \delta \alpha=\alpha \delta,\hspace{3mm} [\beta,\beta^*]=\mu \alpha (\delta-\alpha), \hspace{3mm} [\delta, \beta]=\mu \alpha \beta, \hspace{3mm} \alpha \delta-q^2 \beta^* \beta=1
\end{equation}
where $\mu=1-q^{-2}$. The coproduct and counit $\epsilon$  have the usual matrix coalgebra form. We denote the antipode or 'matrix inverse' by $S$.
\\
We will also work with the dimensional Hopf algebra $\mathcal{A}=B_q[Su_2]$ reduced at $q$ a primitive  $4$'th root of unity. This has  the further relations
\begin{equation}
\beta^4=(\beta^*)^4, \hspace{3mm}\alpha^4=\delta^4=1
\end{equation}
where $\alpha^4,\beta^4,(\beta^*)^4,\delta^4$ generate an undeformed $B_q[SU_2]$ central sub-Hopf algebra of the original $B_q[SU_2]$. Note also that in the reduced case $\delta=\alpha^{-1}(1+q^2 \beta^* \beta)$ is redunant and dim$(\mathcal{A})=4^2$.
in this case a basis of $\mathcal{A}$ is $\{\alpha^p,\beta^r\}$ for $0\le p,r \le 3$. Explicit computations are done via Mathematica for concreteness. Equations involving only the invariant differential forms do not directly involve the function algebra and are solved for all $q^4=1$ and $q^2 \neq 1$.

\section{Exterior algebra}
We take the standard bicovariant exterior algebra on $B_q[SU_2]$ which has the lowest dimensional (4d) space of 1-forms \cite{fuzzy}. Thus, we take a basis $\{e_{\alpha}^{\hspace{3mm}\beta}\}=\left(\begin{array}{cc}
e_a & e_b\\
e_c &e_d \\
   \end{array}
\right)$, where $e_1^{\hspace{3mm}2}=e_b$, etc., and form a right crossed module with right multiplication and the right coaction
\begin{equation}
\Delta_R(e_\alpha^{\hspace{3mm}\beta})=e_\gamma^{\hspace{3mm}\delta} \otimes t_{\hspace{3mm}\alpha}^\gamma St^\beta_{\hspace{3mm}\delta}
\end{equation}
$\Omega^1=\mathcal{A}\otimes\Lambda^1$ is spanned by left-invariant forms as a free left module over $B_q[SU_2]$. These also generate the invariant exterior algebra $\Lambda$ and $\Omega=\mathcal{A}\otimes \Lambda$. Hence $e_a,e_b, e_c$ behave like usual forms or Grassmann variables and
\begin{eqnarray}
\nonumber e_a \wedge e_d+ e_d \wedge e_a+\mu e_c \wedge e_b=0, \hspace{4mm}e_d \wedge e_c+q^2 e_c \wedge e_d +\mu e_a \wedge e_c=0\\
e_b \wedge e_d+q^2 e_d \wedge e_b+\mu e_b \wedge e_a=0, \hspace{4mm} e_d^2=\mu e_c \wedge e_b
\end{eqnarray}
The relations among 1-forms are obtained  by setting to zero the kernel of $\Psi-$id, where $\Psi$ is the crossed-module  braiding (see \cite{fuzzy}). \\
The right module structure on 1-forms is defined via the commutation relations
\begin{eqnarray}
\nonumber [e_a,\alpha ]_q&=&[e_a,\beta ]_{q^{-1}}=[e_c,\beta ]_q=[e_b,\alpha ]_{q^{-1}}=[e_b,\gamma ]_q=0 \\
\nonumber [e_a ,\gamma ]_q&=&\mu \alpha e_b; \hspace{3mm}[e_a,\delta ]_{q^{-1}}=\mu \beta e_b+q\mu^2  \alpha e_a; \hspace{3mm}[e_c,\alpha ]_q=q^2 \mu \beta e_a \\
\nonumber  [e_b,\beta ]_{q^{-1}}&=&\mu \alpha e_a; \hspace{3mm}[e_b,\delta ]_q=q^2 \mu \gamma e_a; \hspace{3mm}[e_d,\alpha ]_{q^{-1}}=\mu \beta e_b \\
\nonumber [e_d,\beta ]_q&=&\mu \alpha e_c+q\mu^2 \beta e_a; \hspace{3mm}[e_d,\gamma ]_{q^{-1}}=\mu(\delta - \alpha )e_b \\
\nonumber  [e_d,\delta ]_q&=&-\mu \beta e_b+q \mu^2 (\delta - \alpha )e_a+\mu \gamma e_c; \hspace{3mm}[e_c,\gamma ]_{q^{-1}}=\mu (\delta - \alpha )e_a+\mu \alpha e_d+q \mu^2 \beta e_b\\
\nonumber  [e_d,\delta ]_{q^{-1}}&=&\mu (q^2-2)\beta e_a+q^2 \mu \beta e_d+q \mu^2 \alpha e_c \\
\label{relationformfunction}
\end{eqnarray}
where $[x,y]_q=xy-qyx$ and $\alpha,\beta,\beta^*,\delta \in B_q[SU_2]$.

\section{Exterior derivative and Lie bracket structure constants}
The differential on the exterior algebra structure is defined by graded anticommutator d$=\mu^{-1}[\theta, \hspace{3mm}\}$ where $\theta=e_a+e_d$. in particular
\begin{equation}
\mathrm{d}e_a=-e_c \wedge e_b; \hspace{5mm}\mathrm{d}e_b=-e_b \wedge (q^{-2} e_a-e_d)\hspace{5mm}\mathrm{d}e_c= e_c \wedge (e_a-q^2 e_d) \hspace{5mm}\mathrm{d}e_d=e_c \wedge e_b
\end{equation}
\begin{lemme}
For all invertible $q^2 \neq 1$
\begin{eqnarray}
\nonumber e_a \alpha^p \beta^r \gamma^s &=& q^{p+s-r} \alpha^p \beta^r \gamma^s e_a+[r]_{q^2} \mu q^{p+s-r-1} \alpha^{p+1} \beta^r \gamma^{s-1} e_b\\
\nonumber e_b \alpha^p \beta^r \gamma^s &=& \mu q^{s-p-r+1}[r]_{q^2} \alpha^{p+1} \beta^{r-1} \gamma^s e_a+(\mu q^{r+s-p-2} [r]_{q^2}[s]_{q^2} \alpha^{p+2} \beta^{r-1} \gamma^{s-1}-q^{s-p-r} \alpha^p \beta^r \gamma^s)e_b \\
\nonumber e_c \alpha^p \beta^r \gamma^s &=& q^{p-r} \alpha^p \beta^r[\mu q^{-1}[s]_{q^2} \gamma (\delta-q^{2(1-s)}\alpha)e_a+(q^{2-r} \mu^2 [s]_{q^2} \beta  \gamma +[s-1]_{q^2} q^{-s} \mu^s \alpha^{s-1} (\delta - \alpha))e_b \\
\nonumber &&+q^{-s}\gamma^s e_c+\mu q^{1-s} [s]_{q^2} \alpha \gamma^{s-1} e_d]\\
\nonumber &&+(\mu q^{p-r+1} [r]_{q^2} \alpha^{p+1} \beta^{r-1} +\mu q^{p+1} [p]_{q^2} \alpha^{p-1} \beta^{r+1})(q^s \gamma^s e_a+\mu q^{s-1} [s]_{q^2} \alpha \gamma^{s-1} e_b)
\\
\nonumber e_d \alpha^p \beta^r \gamma^s &=&q^{2-p-r} \mu^2 [r]_{q^2} \alpha^p \beta^r (1+q^2 [p]_{q^2})(q^s \gamma^s e_a+\mu q^{s-1} [s]_{q^2} \alpha \gamma^{s-1} e_b)+q^{s-r-p+1} \mu [p]_{q^2} \alpha^{p-1} \beta^{r+1} \gamma^s e_b \\
\nonumber &&+q^{r-p-1} \mu [r]_{q^2} \alpha^{p+1} \beta^{r-1} (\mu q^{-1} [s]_{q^2} \gamma (\delta-q^{2(1-s)}\alpha)e_a \\
\nonumber &&+(q^{2-s} \mu^2 [s]_{q^2} \beta \gamma +[s-1]_{q^2} q^{-s} \mu^s \alpha^{s-1} (\delta - \alpha))e_b+q^{-s} \gamma^s e_c+\mu q^{1-s} [s]_{q^2} \alpha \gamma^{s-1} e_d)^\\
 &&+q^{r-p} \alpha^p \beta^r (q^{-s} \gamma^s e_d+q^{1-s} \mu [2]_{q^2} \gamma^{s-1} (\delta - \alpha)e_b)
\end{eqnarray}
where $[n]_{q^2}=\frac{(1-q^{2n})}{(1-q^{2})}$ and the negative powers of $p,r,s$ are omitted.
\end{lemme}
\textit{Proof}
\\
From the commutation relation (\ref{relationformfunction}), we deduce the commutation relations:
\begin{eqnarray}
\nonumber e_a \gamma^s &=&q^s \gamma^s e_a+\mu q^{s-1} [s]_{q^2} \alpha \gamma^{s-1} e_b\\
\nonumber e_b \gamma^s &=&q^s \gamma^s e_b\\
\nonumber e_c \alpha^p &=& q^p \alpha^p e_c+\mu q^{p+1} [p]_{q^2} \alpha^{p-1} \beta e_a\\
\nonumber e_c \gamma^s &=& \mu q^{-1} [s]_{q^2} \gamma (\deg-q^{2(1-s)} \alpha)e_a+(q^{2-s} \mu^2 [s]_{q^2} \beta \gamma +[s-1]_{q^2} q^{-s} \mu^s \alpha^{s-1} (\delta -\alpha))e_b \\
\nonumber && +q^{-s} \gamma^s e_c +\mu q^{1-s} [s]_{q^2} \alpha \gamma^{s-1} e_d\\
\nonumber e_c \beta^r &=& q^{-r}\beta^r e_c +\mu q^{1-r} [r]_{q^2} \alpha \beta^{r-1} e_a\\
\nonumber e_b  \beta^r &=& q^{-r} \beta^r e_b+q^{1-r} \mu [r]_{q^2} \alpha \beta^{r-1} e_a\\
\nonumber e_d \alpha^p &=& q^{-p} \alpha^p e_d +q^{1-p} \mu [p]_{q^2} \alpha^{p-1} \beta e_b\\
\nonumber e_d \beta^r &=& q^r \beta^r e_d+q^{r-1} \mu [r]_{q^2} \alpha \beta^{r-1} e_c+\mu^2 q^{2-r} [r]_{q^2} \beta^r e_a\\
 e_d \gamma^s &=& q^{-s} \gamma^s e_d +q^{1-s}\mu [2]_{q^2} \gamma^{s-1} (\delta -\alpha) e_b
\end{eqnarray} 
These then give the commutation relations with basis elements as stated.
\begin{flushright}
$\spadesuit$
\end{flushright}
From these, we easily obtain
\begin{eqnarray}
\nonumber \mathrm{d}(\alpha^p \beta^r)&=&\mu^{-1}([p]_q q^{-r}(q-1) \alpha^p \beta^r -q^{1-r}[p]_{q^{-1}}(q-1)\mu^2 [r]_{q^2} \alpha^p \beta^r \\
\nonumber &&+\mu^2 q^{-r} [r]_{q^2} \sum_{i=0}^{p-1} [2i+1]_{-q} \alpha^{p-1} \beta^r +[\sum_{i=0}^{r-1} q^{1-i} \mu^i [2i+1]_{-q}+q^{-1} [r]_{q^{-1}}(1-q)\alpha^p \beta^r)e_a\\
\nonumber &&-\mu q^{-1-r} \sum_{i=0}^{p-1} [2i+1]_{-q} \alpha^{p-1} \beta^{r+1} e_b+q^{2r-2-i} \sum_{i=0}^{r-1}[2i+1]_{-q}\alpha^p \beta^{r-1} e_c\\
\nonumber &&+(-\mu^{-1}q^{r-1}(q-1)[p]_{q^{-1}}\alpha^p \beta^r -\mu^{-1} [r]_q (1-q) \alpha^p \beta^{i+1})e_d
\label{derivative}
\end{eqnarray}
obeying the Leibniz rule.
\\
Next, we will need the projection $\tilde{\pi}:\mathcal{A}\to \Lambda^1$ which characterises the above calculus as a quotient of the universal one, i.e. with d$f=f\emptyset \tilde{\pi}(f_{(2)})$ for all  $f\in \mathcal{A}$. Here $\Delta f=f\emptyset \otimes f_{(2)}$ is the Sweedler notation  for the coproduct. Actually $\tilde{\pi}$ can be obtained backwards from d as follows. Let the partial derivatives $\partial^i:\mathcal{A}\to \mathcal{A}$ be defined by d$f=\sum_i (\partial^i f)e_i$. Then 
\begin{equation}
\tilde{\pi}(f)=\sum_i e_i \epsilon (\partial^i f)
\label{pitilde}
\end{equation}
in particular, we obtain
\begin{equation}
\tilde{\pi}(\alpha)=\frac{q}{[2]_q}(qe_a-e_d) \hspace{5mm}\tilde{\pi}(\beta)=e_c \hspace{5mm}\tilde{\pi}(\gamma)=e_b \hspace{5mm}\tilde{\pi}(\delta)=\frac{1}{[2]_q}[q^2e_d-(1+q^{-1})e_a]
\end{equation}
We use the formula (\ref{derivative}) for the exterior derivative. For generic $q$ we compute d$a$ separately. 
\\
Finally, we need the braided-Lie algebra structure constants
\begin{equation}
\mathrm{ad}_R=(\mathrm{id}\otimes \tilde{\pi})\Delta_R, \hspace{4mm}\mathrm{ad}_L=(\tilde{\pi} \otimes \mathrm{id})\delta_L
\end{equation}
where $\Delta_L =(S^{-1} \otimes \mathrm{id})\circ \tau \circ \Delta_R$ is the right coaction converted to a left coaction and $\tau$ is the usual vector space flip. We have
\begin{eqnarray}
\nonumber \mathrm{ad}_R(e_{a})&=& e_c \otimes e_b-q^2 e_d\otimes e_b +(\nu e_a+\xi 
e_d)\otimes e_b\\
\nonumber  \mathrm{ad}_R (e_{d})&=& -q^2 e_a\otimes e_c+e_b \otimes e_c -\lambda (\nu 
e_a+\xi e_d)\otimes e_c\\
\nonumber  \mathrm{ad}_R (e_{b})&=& \frac{q}{[2]_q}e_b\otimes (q e_a-e_d)-q^2 e_c\otimes 
e_b +e_d\otimes e_b\\
\nonumber  \mathrm{ad}_R (e_{c})&=& e_a\otimes e_c-q^2 e_b\otimes e_c +\frac{1}
{[2]_q}e_c\otimes (q^2 e_d-(1+q^{-1})e_a)\\
\nonumber  \mathrm{ad}_L (e_{a})&=& -q^2 e_b \otimes e_c -q^2 e_b \otimes e_d +e_b \otimes (\nu e_a+\xi e_d)\\
\nonumber  \mathrm{ad}_L (e_{d})&=& -q^2 e_c \otimes e_a -q^2 e_c\otimes e_b +e_c\otimes \lambda (\nu e_a +\xi e_d)\\
\nonumber  \mathrm{ad}_L (e_{b})&=& -q^2 e_b \otimes e_d -q^2 e_b \otimes e_c +\frac{1}{[2]_q}(q^2 e_d-(1+q^{-1})e_a)\otimes e_b\\
  \mathrm{ad}_L (e_{c})&=& -q^2 e_c \otimes e_a-q^2 e_c \otimes e_b +\frac{q}{[2]_q}(q e_a-e_d)\otimes e_c
  \label{ad}
\end{eqnarray}
where 
\begin{eqnarray}
\nonumber \nu &=&\frac{q^3-24q^2+7q+34}{5q^3-15q^2-50q-30}=\frac{q^2 A_d^a}{(A_d^a)^2-(A_a^a)^2}\\
\nonumber \xi &=&\frac{q^3-15q-13}{5q^3-15q^2-50q-30}=\frac{q^2 A^a_a}{-(A_d^a)^2+(A_a^a)^2}\\
\lambda&=& \frac{1+2q+2q^2}{-2+2q^2+q^3}
\end{eqnarray}
We define by $\mathrm{ad}(e_i)=\sum_{j,k} \mathrm{ad}(jk|i)e_j\otimes e_k$, where we use the indices $i,j,k$ to run through $\{e_a,e_b,e_c,e_d\}$.

\section{Spin connection and Riemannian curvature}
Still with $q$ arbitrary, there is a natural $\Delta_R$-covariant Killing metric in \cite{riemanngeometry} of the form
\begin{equation}
\eta = e_c \otimes e_b+q^2 e_b \otimes e_c+\frac{e_a\otimes e_a -q e_a\otimes e_d -q e_d\otimes e_a +q(q^2+q-1) e_d\otimes e_d}{[2]_q}+\rho \theta \otimes \theta
\end{equation}
where $\rho =\frac{q(1-q-q^2)}{1+q}$ is the natural choice for the Hodge * operator as explained in \cite{CqSL2}.This is the local cotangent space of $\C_q[SL_2]$ with $\theta$ an intrinsic 'time' direction induced by noncommutative geometry. We can add any multiple of $\theta \otimes \theta$ and still retain $\Delta_R$-invariance.\\
For any such invariant metric, we have symmetry in the sense 
\begin{equation}
\wedge (\eta)=0
\end{equation}
Moreover, the equations for a torsion-free and skew-metric-compatible 'generalised Levi-Civita' spin connection become independent  of $\eta$ and reduce to the torsion and 'cotorsion' equations \cite{riemanngeometry}:
\begin{equation}
\mathrm{d}e_i+\sum_{j,k} A_j \wedge e_k \mathrm{ad}_L (jk|i)=0, \hspace{3mm}\mathrm{d}e_i+\sum_{j,k} e_j \wedge A_k \mathrm{ad}_R(jk|i)=0
\label{torsioncotorsioneq}
\end{equation}
in these equations we write $A(e_i) \equiv A_i$, and a generalised spin connection is given by four such forms $A_a, A_b, A_c, A_d$ obeying (\ref{dei}). in principle there is also an optional 'regularity' condition as explained in \cite{riemanngeometry} which ensures that the curvature is braided-Lie algebra valued. By the same arguments as above, this regularity condition can be written as
\begin{equation}
\sum_{i,j} A_i \wedge A_j  \epsilon(\partial^i \partial^j  f)=0, \hspace{3mm} \forall f \in \ker \tilde{\pi}
\end{equation}
\begin{theorem}
For generic $q$ or for $q$ an $4$-th root of unity , if we define $A_i=A^{\hspace{3mm}j}_{i}e_j$, there is a unique torsion-free and cotorsion free spin connection given by
\begin{eqnarray}
\nonumber 
A^{\hspace{3mm}d}_{d}&=&A^{\hspace{3mm}a}_{a}=\frac{4+6q+5q^2+3q^3}
{5-4q^2+2q^3}\\
\nonumber 
A^{\hspace{3mm}a}_{d}&=&A^{\hspace{3mm}d}_{a}=\frac{-1+5q+7q^2+q^3}
{2+5q-4q^3}\\
\nonumber 
A^{\hspace{3mm}b}_{d}&=&-\frac{275+120q-140q^2+15q^3}
{-+106q+14q^2-84q^3}\\
\nonumber 
A^{\hspace{3mm}c}_{d}&=&-\frac{295-655q-430q^2+48q^3}
{319-112q-333q^2+99q^3}\\
\nonumber 
A^{\hspace{3mm}b}_{a}&=&-\frac{146+270q+25q^2-90q^3}
{14-84q+9q^2+106q^3}\\
\nonumber 
A^{\hspace{3mm}c}_{a}&=&-\frac{330-285q-465q^2+292q^3}
{99+319q-112q^2-333q^3}\\
\nonumber 
A^{\hspace{3mm}c}_{c}&=&\frac{1+2q+2q^2}
{-2+2q^2+q^3}\\
A^{\hspace{3mm}b}_{b}&=&\frac{-2-3q+3q^2+2q^3}
{2+5q-4q^3}
\label{connection}
\end{eqnarray} 
The connection is not in general regular.
\end{theorem}
\textit{Proof}
\\
Looking first at the torsion equation, we see that the coefficients of $A$ are all to the left and hence its functional dependence is immaterial.
We write out the equations using the form of $\mathrm{ad}_R$ and d$e_a$ from (\ref{ad}) and match coefficients of a absis of $\Lambda^2$. This is a linear system:
\begin{eqnarray}
\nonumber -e_c \wedge e_b+e_c \wedge A_b-q^2 e_d \wedge A_b+(\nu A_a +\xi A_d)\wedge e_b &=&0\\
\nonumber e_c \wedge e_b -q^2 e_a A_c+e_b \wedge A_c +\lambda (\nu A_a +\xi A_d)\wedge e_c&=&0\\
\nonumber -q^{-2} e_b \wedge e_a +e_b \wedge e_d -\frac{q}{[2]_q}(e_b \wedge (A_d - qA_a))-q^2 e_c \wedge A_b +e_d \wedge A_b &=&0\\
 -q^{-2} e_c \wedge e_a+e_c \wedge e_d +e_a \wedge A_c -q^2 e_b \wedge A_c +\frac{q^2}{[2]_q}e_c \wedge A_d-\frac{(1+q^{-1})}{[2]_q}(e_c \wedge A_a&=&0
 \label{torsionsysteq}
\end{eqnarray}
Because the relations between two forms
\begin{eqnarray}
\nonumber e_a \wedge e_d +e_d \wedge e_a +\mu e_c \wedge e_b &=&0\\
\nonumber e_d \wedge e_c +q^2 e_c \wedge e_d +\mu e_a \wedge e_c&=&0\\
\nonumber e_b \wedge e_d+q^2 e_d \wedge e_b +\mu e_b \wedge e_a&=&0\\
\nonumber e_d^2-\mu e_c \wedge e_b&=&0\\
\nonumber e_d \wedge e_a +e_a \wedge e_d +\mu e_b \wedge e_c &=&0\\
\nonumber e_c \wedge e_d +q^2 e_d \wedge e_c +\mu e_c \wedge e_a&=&0\\
\nonumber e_d \wedge e_b+q^2 e_b \wedge e_d +\mu e_a \wedge e_b&=&0\\
\end{eqnarray}
and (\ref{torsionsysteq}) we obtain the equations
\begin{eqnarray}
\nonumber -\mu (1+q^{-2})-\nu A^{\hspace{3mm}a}_{a}-\xi A^{\hspace{3mm}a}_{d}&=&0\\
\nonumber q^{-2}-\nu A^{\hspace{3mm}d}_{a}-\xi A^{\hspace{3mm}d}_{d}&=&0\\
\nonumber -1+A^{\hspace{3mm}b}_{b}- \nu A^{\hspace{3mm}c}_{a}-\xi A^{\hspace{3mm}c}_{d})&=&0\\
\nonumber A^{\hspace{3mm}c}_{c}-q^{-2} \lambda (\nu A^{\hspace{3mm}a}_{a}-\xi A^{\hspace{3mm}a}_{d})&=&0\\
\nonumber -1-A^{\hspace{3mm}c}_{c}-\lambda (\nu A^{\hspace{3mm}b}_{a}+\xi A^{\hspace{3mm}b}_{b})\\
\nonumber \nu A^{\hspace{3mm}d}_{a}+\xi A^{\hspace{3mm}d}_{d}&=&0\\
\nonumber -1-\frac{q^3}{[2]_q}(A^{\hspace{3mm}a}_{d}-qA^{\hspace{3mm}a}_{a})-\mu A^{\hspace{3mm}b}_{b}&=&0\\
\nonumber 1-\frac{q}{[2]_q}(A^{\hspace{3mm}d}_{d}-qA^{\hspace{3mm}d}_{a})-q^2 A^{\hspace{3mm}b}_{b}&=&0\\
\nonumber \frac{q}{[2]_q}(A^{\hspace{3mm}c}_{d}-q A^{\hspace{3mm}c}_{a})-q^2 A^{\hspace{3mm}b}_{b}+A^{\hspace{3mm}a}_{b}&=&0\\
\nonumber -q^2 A^{\hspace{3mm}a}_{b}-\mu A^{\hspace{3mm}c}_{b}&=&0\\
\nonumber A^{\hspace{3mm}d}_{b}-q^2 A^{\hspace{3mm}c}_{b}&=&0\\
\nonumber  A^{\hspace{3mm}a}_{b}=A^{\hspace{3mm}c}_{b}=A^{\hspace{3mm}d}_{b}&=&0\\
\nonumber -1+A^{\hspace{3mm}c}_{c}+\frac{1}{[2]_q}A^{\hspace{3mm}a}_{d}-q^2 \frac{(1+q^{-1})}{[2]_q}A^{\hspace{3mm}a}_{a}&=&0\\
\nonumber 1+\frac{q^2}{[2]_q} A^{\hspace{3mm}d}_{d}-
\frac{(1+q^{-1})}{[2]_q}A^{\hspace{3mm}d}_{a}&=&0\\
\nonumber A^{\hspace{3mm}b}_{c}-A^{\hspace{3mm}a}_{c}&=&0\\
\nonumber A^{\hspace{3mm}d}_{c}&=&0\\
\nonumber -q^2 A^{\hspace{3mm}c}_{c}+\frac{q^2}{[2]_q} A^{\hspace{3mm}b}_{d}-\frac{(1+q^{-1})}{[2]_q}A^{\hspace{3mm}b}_{a}&=&0
\end{eqnarray} 
With the help of mathematica, we find the only solution of this equations system which is (\ref{connection}). 
\begin{flushright}
$\spadesuit$
\end{flushright}

The covariant derivative $\Omega^1 \to \Omega^1 \otimes_{\mathcal{A}} \Omega^1$ is comuted from \cite{riemanngeometry}
\begin{equation}
\nabla e_i=-\sum_{j,k} A_j \otimes e_k \mathrm{ad}_L(jk|i), \hspace{4mm} \forall i
\end{equation}
it obeys the usual  derivation-like rule for covariant derivatives, so we need only give it on basic 1-forms. For the above canonical spin connection it comes out (in a similar manner to solving the torsion equation in 
Theorem 1) as 
\begin{eqnarray}
\nonumber \nabla e_a &=&A_b \otimes (q^2 e_c +q^2 e_d-\nu e_a -\xi e_d)\\
\nonumber \nabla e_d &=&A_c \otimes (q^2 e_a+q^2 e_b+ \lambda(\nu e_a +\xi e_d))\\
\nonumber \nabla e_b &=&A_b \otimes (q^2 e_d +q^2 e_c)-\frac{1}{[2]_q}(q^2 A_d +(1+q^{-1})A_a)\otimes e_b\\
 \nabla e_c&=& A_c\otimes (q^2 e_a+q^2 e_b)-\frac{q}{[2]_q}(qA_a-A_c)\otimes e_c
 \label{nabla}
\end{eqnarray}
Finally, for  any connection, the Riemannian curvature is computed from 
\cite{riemanngeometry}
\begin{equation}
\mathrm{Riemann}=((\mathrm{id}\wedge \nabla )-\mathrm{d}\otimes \mathrm{id})\circ \nabla 
\end{equation}
or equally well from the curvature $F=\mathrm{d}A+A*A : \ker \epsilon \to \Omega^2$ of $A$. When the connection is not regular, the latter Yang-Mills curvature does not descend to a map $\Lambda^1 \to \Omega^2$ (it is not  'Lie algebra valued but lives in the enveloping algebra of the braided-Lie algebra). However, this does not directly affect the Riemannian geometry (it merely  complicates the geometry 'upstairs' on the quantum frame bundle); in the proof of \cite{riemanngeometry} Corol 3.8 one should simply omit the $\tilde{\pi}$ in the argument of $F$ for the relation to the Riemann curvature.
\begin{proposition}
The Riemann curvature $\Omega^1 \to \Omega^2 \otimes_{\mathcal{A}} \Omega^1$ of the canonical spin connection in Theorem 1 is 
\begin{eqnarray}
\nonumber \mathrm{Riemann}(e_a)&=&A_b \wedge A_c \otimes (e_a+e_b)-\frac{q^3}{[2]_q}A_b \wedge (qA_a-A_c)\otimes e_c +q^2 \lambda A_b \wedge A_c \otimes (\nu e_a+ \xi e_d)\\
\nonumber  \mathrm{Riemann}(e_b)&=&q^2 A_b \wedge (A_c \otimes (q^2 e_a +q^2 e_b +\lambda (\nu e_a +\xi e_d))-\frac{q}{[2]_q}(q A_a-A_c)\otimes e_c)\\
\nonumber \mathrm{Riemann}(e_c)&=&q^2 A_c \wedge (A_b \otimes (q^2 e_c +q^2 e_d -\nu e_a -\xi e_d)+A_b \otimes (q^2 e_d+q^2 e_c)-\frac{1}{[2]_q}(q^2 A_d-(1+q^{-1})A_a)\otimes e_b)\\
\nonumber \mathrm{Riemann}(e_d)&=&A_c \wedge A_b \otimes (e_c+e_d)-q^2 A_c \wedge A_b \otimes (\nu e_a +\xi e_d)-\frac{q^2}{[2]_q}A_c \wedge (q^2 A_d -(1+q^{-1})A_a)\otimes e_b
\end{eqnarray}
\end{proposition}
\textit{Proof}
Direct computation using the relations in the preliminaries and the formula
(\ref{nabla}) for $\nabla$. Note that Riemann is a tensor, so that Riemann$(fe_a)=f$Riemann$(e_a)$ for all $f \in \mathcal{A}$, i.e. we need  only give it on the basic 1-form.
\begin{flushright}
$\spadesuit$
\end{flushright}

\section{Dirac operator}
The additional ingredient for a Dirac operator is a choice of spinor representation $W$ of the frame quantum group and equivariant gamma-matrices $\gamma : \Lambda^1 \to \mathrm{End}(W)$. The spinor bundle in our case is just the tensor product $\mathcal{A}\otimes W$, which is the space of spinors. We take the 2-dimensional representation (i.e. a Weyl spinor) so a spinor  has components $\psi^\alpha \in B_q[SU_2]$ for $\alpha =1,2$
\paragraph{}
Since $\Lambda^1$ for a differential calculus of $B_q[SU_2]$ was originally given in the endomorphism basis $\{e_\alpha^{\hspace{2mm}\beta} \}$, the canonical gamma-matrices proposed in \cite{riemanngeometry} are just the identity map in that basis. Or in terms of our above $\{e_i\}$ they provide the conversion according to
\begin{equation}
\gamma (e_i)^{\alpha}_{\hspace{2mm}\beta} e_{\alpha}^{\hspace{2mm}\beta}=e_i, \hspace{3mm} 
\gamma(e_a)=\left(\begin{array}{cc}
1&0\\
0&0\\ 
   \end{array}
\right),\hspace{3mm}
\gamma(e_a)=\left(\begin{array}{cc}
0&1\\
0&0\\ 
   \end{array}
\right), etc.
\end{equation}
if we take more usual linear combinations $e_x, e_y ,e_z, \theta$ (where $e_x, e_y$ are linear combinations of $e_b,e_c$) then the gamma-matrices would have a  more usual form of Pauli matrices and the identity, but this is not particularly natural when $q \neq 1$ given that our metric is not symmetric.
\paragraph{}
Using the $\{e_\alpha^{\hspace{2mm}\beta}$ basis, the canonical Dirac operator in \cite{riemanngeometry} is 
\begin{equation}
(\cancel{D} \psi)^{\alpha}=\partial^{\alpha}_{\hspace{2mm} \beta}   \psi^{\beta}-A(\tilde{\pi}S^{-1}) t^{\gamma}_{\hspace{2mm}\beta}  )^{\alpha}_{\hspace{2mm}\gamma} \psi^{\beta}
\label{dirac}
\end{equation}
where 
\begin{equation}
A=A^\alpha_{\hspace{2mm}\beta} e_\alpha^{\hspace{2mm}\beta}, \hspace{3mm} \mathrm{d}f=\partial^{\alpha}_{\hspace{2mm}\beta}(f)e_{\alpha}^{\hspace{2mm}\beta}, \hspace{3mm} \forall A \in \Omega^1, \hspace{3mm} f \in B_q[SU_2]
\end{equation}
\begin{proposition}
The Dirac operator for the canonical spin connection on $B_q[SU_2]$ in Theorem 1 is
\begin{equation}
\cancel{D}=\left(\begin{array}{cc}
\partial^a -\frac{1+q^{-1}-q}{1+q}A_d^{\hspace{3mm}a}+A_a^{\hspace{3mm}a}+q^2 A_b^{\hspace{3mm}b} &\partial^b -\frac{1+q^{-1}-q}{1+q}A_d^{\hspace{3mm}c}+A_a^{\hspace{3mm}c} \\
\partial^c -\frac{q^2}{[2]_q}A_a^{\hspace{3mm}b}+\frac{q}{[2]_q}A_d^{\hspace{3mm}b} &\partial^d-\frac{q^2}{[2]_q}A_a^{\hspace{3mm}d}+\frac{q}{[2]_q}A_d^{\hspace{3mm}d}+q^2A_c^{\hspace{3mm}c} \\ 
   \end{array}
\right)
\end{equation}
\end{proposition}
\textit{Proof}
\\
From  (\ref{pitilde}) for $\tilde{\pi}$, and the specific form of our spin connection from Theorem 1, we have
\begin{eqnarray}
\nonumber A(\tilde{\pi} S^{-1} \alpha)&=&-A_a+\frac{1+q^{-1}-q}{1+q}A_d\\
\nonumber A(\tilde{\pi} S^{-1} \beta)&=&-q^2 A_c\\
\nonumber A(\tilde{\pi} S^{-1} \gamma)&=&-q^2 A_b\\
 A(\tilde{\pi} S^{-1} \delta)&=&\frac{q^2}{[2]_q}A_a-\frac{q}{[2]_q}A_d
\end{eqnarray}
We then convert to the spinor basis with $(e_i)^{\alpha}_{\hspace{2mm}\beta}=\gamma(e_i)^{\alpha}_{\hspace{2mm}\beta}$ so, e.g. $(e_b)^1_{\hspace{2mm}2}=1$ and its other components are zero. We then compute the matrix $\cancel{A}^{\alpha}_{\hspace{2mm}\beta}=A(\tilde{\pi}S^{-1} t^{\gamma}_{\hspace{2mm}\beta})^{\hspace{2mm} \alpha}_{\gamma}$ so that $\cancel{D}=\cancel{\partial}-\cancel{A}$. We have 
\begin{eqnarray}
\nonumber \cancel{A}^1_{\hspace{2mm}1}&=&\frac{1+q^{-1}-q}{1+q}A^{\hspace{2mm}a}_d -A^{\hspace{2mm}a}_a-q^2 A^{\hspace{2mm}b}_b\\
\nonumber  \cancel{A}^2_{\hspace{2mm}1}&=& \frac{1+q^{-1}-q}{1+q}A^{\hspace{2mm}c}_d -A^{\hspace{2mm}c}_a\\
\nonumber \cancel{A}^1_{\hspace{2mm}2}&=& \frac{q^2}{[2]_q}A^{\hspace{2mm}b}_a -\frac{q}{[2]_q}A^{\hspace{2mm}b}_d\\ \cancel{A}^2_{\hspace{2mm}2}&=& \frac{q^2}{[2]_q}A^{\hspace{2mm}d}_a -\frac{q}{[2]_q}A^{\hspace{2mm}d}_d-q^2 A^{\hspace{2mm}c}_c
\end{eqnarray}
which gives the results as stated
\begin{flushright}
$\spadesuit$
\end{flushright}
We see that the Dirac operator is not the naive $\partial$ that one might write guess without a spin connection. This is not the same phenomenon as for $S_3$ in \cite{riemanngeometry}.
\begin{proposition}
For $q=1$ the approximation of the eigenvalues of the Dirac operator (\ref{dirac}) are  given
by
\begin{eqnarray}
\nonumber && 6.13535, 6.09138 - 0.0458136 i, 6.09138 + 0.0458136 i, 6.04356, 
 5.69131 + 1.88396 i, 5.69131 - 1.88396 i,\\
 \nonumber  && 4.92273 + 2.56606 i, 
 4.92273 - 2.56606 i, 5.44035 + 0.691008 i, 
 5.44035 - 0.691008 i, 4.69388,\\
 \nonumber  && 4.2556 - 1.54516 i, 
 4.2556 + 1.54516 i, 4.36549 + 0.354504 i, 4.36549 - 0.354504 i, 3.63451 + 2.3545 i, \\
 \nonumber  &&  3.63451 - 2.3545 i, 3.95644, 
 3.90862 - 0.0458136 i, 3.90862 + 0.0458136 i, 3.86465, 
 3. + 2.11925 i, 3. - 2.11925 i, \\
 \nonumber &&3.07727 + 0.566058 i, 
 3.07727 - 0.566058 i, 2.55965 + 1.30899 i, 2.55965 - 1.30899 i, \\
 \nonumber
&& 1.7444 - 1.54516 i, 1.7444 + 1.54516 i, 2.30869 + 0.116035 i, 
 2.30869 - 0.116035 i, 1.30612 
\end{eqnarray}
for $q=i$the approximation of the eigenvalues of the Dirac operator (\ref{dirac}) are  given by
\begin{eqnarray}
\nonumber &&-4.96224 + 0.188313 i, -4.94028 + 0.196129 i, -4.86666 + 
 0.199838 i, -4.81671 + 0.199838 i,\\
 \nonumber &&-3.8887 + 1.36007 i, -4.0635 + 
 0.0761453 i, -3.84125 + 1.26724 i, -3.7927 - 0.921684 i, -3.86394 + 
 0.0590435 i,
 \\
 \nonumber &&  -3.76238 - 0.838047 i, -3.28813 + 
 0.320025 i, -2.97248 + 0.288083 i, 1.20446 + 2.56874 i, 2.37401 + 
 1.45875 i,\\
 \nonumber && 2.35204 + 1.45093 i, 1.17415 + 2.48511 i, 2.27843 + 
 1.44722 i, -2.65153 + 0.260679 i, 2.22847 + 1.44722 i, \\
 \nonumber && -2.56214 + 
 0.239703 i, -2.36421 + 0.187262 i, -2.29997 + 0.184081 i, 1.47527 + 
 1.57091 i, 1.2757 + 1.58802 i,
  \\
 \nonumber &&  0.699899 + 1.32703 i, -0.288263 + 
 1.46298 i, -0.224021 + 1.4598 i, 0.384248 + 1.35898 i, -0.0260945 + 
 1.40736 i, \\
 \nonumber &&  0.0632908 + 1.38638 i, 1.30046 + 0.286992 i, 1.25301 + 
 0.379819 i
 \label{i}
\end{eqnarray}
for $q=-i$the approximation of the eigenvalues of the Dirac operator (\ref{dirac}) are  given by the complex conjugate of (\ref{i}) 
\begin{eqnarray}
\nonumber && -4.96224 - 0.188313 i, -4.94028 - 0.196129 i, -4.86666 - 
 0.199838 i, -4.81671 - 0.199838 i, \\
 \nonumber &&  -3.8887 - 1.36007 i,-4.0635 - 
 0.0761453 i, -3.84125 - 1.26724 i, -3.7927 + 0.921684 i, -3.86394 - 
 0.0590435 i, \\
 \nonumber &&  -3.76238 + 0.838047 i, -3.28813 - 
 0.320025 i, -2.97248 - 0.288083 i, 1.20446 - 2.56874 i, 2.37401 - 
 1.45875 i, \\
 \nonumber &&  2.35204 - 1.45093 i, 1.17415 - 2.48511 i, 2.27843 - 
 1.44722 i, -2.65153 - 0.260679 i, 2.22847 - 1.44722 i, \\
 \nonumber &&  -2.56214 - 
 0.239703 i, -2.36421 - 0.187262 i, -2.29997 - 0.184081 i, 1.47527 - 
 1.57091 i, 1.2757 - 1.58802 i, \\
 \nonumber &&  0.699899 - 1.32703 i, -0.288263 - 
 1.46298 i, -0.224021 - 1.4598 i, 0.384248 - 1.35898 i, -0.0260945 - 
 1.40736 i, \\
 \nonumber &&  0.0632908 - 1.38638 i, 1.30046 - 0.286992 i, 1.25301 - 
 0.379819 i
\end{eqnarray}
\end{proposition}
\textit{Proof}
\\
It is easiest (and most natural mathematically) to compute the unnormalised $\cancel{\partial}$ where $\mathrm{d}=[\theta, \}$ without the factor $\mu^{-1}$ that was inserted for the classical limit. 
In the basis
\begin{equation}
1,\beta,\beta^2,\beta^3,\alpha, \alpha \beta,\alpha \beta^2,\alpha \beta^3, \alpha^2,\alpha^2 \beta,\alpha^2 \beta^2,\alpha^2 \beta^3,\alpha^3, \alpha^3 \beta, \alpha^3 \beta^2, \alpha^3 \beta^3
\end{equation}
we can calculate the right translation operator of $\alpha$:
\begin{eqnarray}
\nonumber R_{\alpha}&=&
\left(\begin{array}{cccccccccccccccc}
0&0&0&0&0&0&0&0&0&0&0&0&1&0&0&0\\
0&0&0&0&0&0&0&0&0&0&0&0&0&q^2&0&0\\
0&0&0&0&0&0&0&0&0&0&0&0&0&0&1&0\\
0&0&0&0&0&0&0&0&0&0&0&0&0&0&0&q^2\\
1&0&0&0&0&0&0&0&0&0&0&0&0&0&0&0\\
0&q^2&0&0&0&0&0&0&0&0&0&0&0&0&0&0\\ 
0&0&1&0&0&0&0&0&0&0&0&0&0&0&0&0\\
0&0&0&q^2&0&0&0&0&0&0&0&0&0&0&0&0\\
0&0&0&0&1&0&0&0&0&0&0&0&0&0&0&0\\
0&0&0&0&0&q^2&0&0&0&0&0&0&0&0&0&0\\
0&0&0&0&0&0&1&0&0&0&0&0&0&0&0&0\\
0&0&0&0&0&0&0&q^2&0&0&0&0&0&0&0&0\\  
0&0&0&0&0&0&0&0&1&0&0&0&0&0&0&0\\  
0&0&0&0&0&0&0&0&0&q^2&0&0&0&0&0&0\\
0&0&0&0&0&0&0&0&0&0&1&0&0&0&0&0\\
0&0&0&0&0&0&0&0&0&0&0&q^2&0&0&0&0\\
 \end{array}
\right)
\end{eqnarray}
of $\beta$
\begin{eqnarray}
\nonumber R_{\beta}&=&
\left(\begin{array}{cccccccccccccccc}
0&0&0&1&0&0&0&0&0&0&0&0&0&0&0&0\\
1&0&0&0&0&0&0&0&0&0&0&0&0&0&0&0\\
0&1&0&0&0&0&0&0&0&0&0&0&0&0&0&0\\
0&0&1&0&0&0&0&0&0&0&0&0&0&0&0&0\\
0&0&0&0&0&0&0&1&0&0&0&0&0&0&0&0\\
0&0&0&0&1&0&0&0&0&0&0&0&0&0&0&0\\ 
0&0&0&0&0&1&0&0&0&0&0&0&0&0&0&0\\
0&0&0&0&0&0&1&0&0&0&0&0&0&0&0&0\\
0&0&0&0&0&0&0&0&0&0&0&1&0&0&0&0\\
0&0&0&0&0&0&0&0&1&0&0&0&0&0&0&0\\
0&0&0&0&0&0&0&0&0&1&0&0&0&0&0&0\\
0&0&0&0&0&0&0&0&0&0&1&0&0&0&0&0\\  
0&0&0&0&0&0&0&0&0&0&0&0&0&0&0&1\\  
0&0&0&0&0&0&0&0&0&0&0&0&1&0&0&0\\
0&0&0&0&0&0&0&0&0&0&0&0&0&1&0&0\\
0&0&0&0&0&0&0&0&0&0&0&0&0&0&1&0\\
 \end{array}
\right)
\end{eqnarray}
of $\beta^*$
\begin{eqnarray}
\nonumber 
R_{\beta^*}&=&
\left(\begin{array}{cccccccccccccccc}
0&-q^2&0&0&0&0&0&0&0&0&0&0&0&1&0&0\\
0&0&-q^2&0&0&0&0&0&0&0&0&0&0&0&q^2&0\\
0&0&0&-q^2&0&0&0&0&0&0&0&0&0&0&0&1\\
-q^2&0&0&0&0&0&0&0&0&0&0&0&q^2&0&0&0\\
0&1&0&0&0&-q^2&0&0&0&0&0&0&0&0&0&0\\
0&0&q^2&0&0&0&-q^2&0&0&0&0&0&0&0&0&0\\ 
0&0&0&1&0&0&0&-q^2&0&0&0&0&0&0&0&0\\
q^2&0&0&0&-q^2&0&0&0&0&0&0&0&0&0&0&0\\
0&0&0&0&0&1&0&0&0&-q^2&0&0&0&0&0&0\\
0&0&0&0&0&0&q^2&0&0&0&-q^2&0&0&0&0&0\\
0&0&0&0&0&0&0&1&0&0&0&-q^2&0&0&0&0\\
0&0&0&0&q^2&0&0&0&-q^2&0&0&0&0&0&0&0\\  
0&0&0&0&0&0&0&0&0&1&0&0&0&-q^2&0&0\\  
0&0&0&0&0&0&0&0&0&0&q^2&0&0&0&-q^2&0\\
0&0&0&0&0&0&0&0&0&0&0&1&0&0&0&-q^2\\
0&0&0&0&0&0&0&0&q^2&0&0&0&-q^2&0&0&0\\
 \end{array}
\right)
\end{eqnarray}
and for $\delta$
\begin{eqnarray}
R_{\alpha}&=&R_\delta
\end{eqnarray}
From the definition of the partial derivative 
\begin{equation}
\partial^a=R_a-\mathrm{id} 
\end{equation}
we can totally explicit the Dirac operator (\ref{dirac}).
With the help of mathematica, we calculate the eigenvalues of this Dirac operator which are the values given in the proposition.
\begin{flushright}
$\spadesuit$
\end{flushright}

\newpage

\section{Discussion}
We use the powerful software Mathematica to solve the torsion cotorsion equation (\ref{torsioncotorsioneq}). We find a  difficult solution for the Levi-Civita connection $A$ because $A_a$ and $A_d$ have 4 components on the basis of 1-forms. 
\\
When we tried to compute the lifting and the Ricci curvature as in \cite{CQSL2}, we need $\Psi$ but we didn't know how to calculate it.
\\
This gives rise to a very difficult Dirac operator (\ref{dirac}). This case is very different of the case of $D_6$ given in \cite{D6} because it is not  the  naive $\partial$ without a spin connection. 
We try to give an exact expression for the eigenvalues but the calculus time are too long and Mathematica was not able to finish it in a finit time. 
\\
Finally, because we didn't know exactly the expression of the eigenvalues. We were not able to compute the eigenmodes of the Dirac operator found previously.

\end{document}